\documentclass[10pt,a4paper,twoside]{article}
\usepackage{epsfig}
\usepackage{array}
\usepackage{longtable}
\usepackage{baltlat5}
\usepackage{wrapfig}
\begin{document}
\ \ \vspace{-0.5mm}

\setcounter{page}{547} \vspace{-2mm}

\titlehead{Baltic Astronomy, vol.\ts 15, 547--560, 2006.}

 \titleb{A SURVEY OF COMPACT STAR CLUSTERS IN THE SOUTH-WEST FIELD OF
THE M\,31 DISK. STRUCTURAL PARAMETERS}

\begin{authorl}
\authorb{I.~\v{S}ablevi\v{c}i\={u}t\.{e}}{1},
\authorb{V.~Vansevi\v{c}ius}{1},
\authorb{K.~Kodaira}{2},
\authorb{D.~Narbutis}{1},
\authorb{R.~Stonkut\.{e}}{1,3} and
\authorb{A.~Brid\v{z}ius}{1}
\end{authorl}

\begin{addressl}
\addressb{1}{Institute of Physics, Savanori\c{u} 231,
Vilnius LT-02300, Lithuania \\ wladas@astro.lt}
\addressb{2}{The Graduate University for Advanced Studies (SOKENDAI), \\
Shonan Village, Hayama, Kanagawa 240-0193, Japan }
\addressb{3}{Vilnius University Observatory, \v{C}iurlionio 29, Vilnius LT-03100, Lithuania}
\end{addressl}

\submitb{Received 2006 November 30; accepted 2006 December 21}

\begin{summary}
We present structural parameters for 51 compact star clusters from
the survey of star clusters conducted in the South-West field of
the M\,31 disk by Kodaira et al. (2004). Structural parameters
of the clusters were derived by fitting the 2-D King and EFF (Elson,
Fall and Freeman 1987) models to the $V$-band cluster images.
Structural parameters derived for two M\,31 clusters, which are in
common with the study based on the HST data (Barmby et al. 2002),
are consistent with earlier determination. The M\,31 star cluster
structural parameters in general are compatible with the
corresponding Milky Way galaxy and Magellanic Clouds cluster
parameters.

\end{summary}


\resthead{Compact star clusters in the M\,31
disk}{I.~\v{S}ablevi\v{c}i\={u}t\.{e}, V.~Vansevi\v{c}ius,
K.~Kodaira et al.}

\sectionb{1}{INTRODUCTION}

The study of the formation and evolution of star clusters has been
proved to be a valuable tool for disentangling star formation history of
the host galaxy disks (Efremov \& Elmegreen 1998).  Star clusters may
survive long enough under dynamics of the galactic disk, thus preserving
the imprint of the conditions of star-forming events that generated
them.  Therefore, a rigorous estimation of the evolutionary status,
chemical composition and kinematics of the clusters is vital for
understanding of the structure and formation of galaxies in general
(Williams \& Hodge 2001b; Morrison et al. 2004; Burstein et al. 2004).
A growing interest in this field is showing up in different studies of
star cluster systems in various galaxies, see e.g., Small Magellanic
Cloud (Rafelski \& Zaritsky 2005; Hill \& Zaritsky 2006) and M\,51
(Bastian et al. 2005; Lee et al. 2005).

The imaging with Suprime-Cam (Miyazaki et al. 2002) at the Subaru
Telescope (National Astronomical Observatory of Japan) provides a unique
opportunity for a detailed study of star clusters in the Local Group
galaxies.  In this context M\,31 surpasses other Local Group galaxies in
a number of clusters (Galleti et al. 2004) and it is suitable for a
study of compact stellar clusters in the galaxy disk (Kodaira et al.
2004; hereafter Paper I).  The results of extensive surveys and studies
of the M\,31 cluster system became available recently (Galleti et al.
2006; Fan et al. 2006; Beasley et al. 2005; Fusi Pecci et al. 2005;
Galleti et al. 2004; Beasley et al. 2004; Barmby et al. 2002; Williams
\& Hodge 2001a).

In this study we present basic structural parameters derived for 51
compact star clusters taken from the survey performed in the S-W field
of the M\,31 disk by Kodaira et al.  (2004).  The high resolution
Suprime-Cam imaging enabled us to resolve a large fraction of star
clusters in M\,31.  Therefore, the sample of the clusters selected for
this study consists of the unresolved, semi-resolved and well resolved
objects.  We employed the widely used 2-D profile fitting program
package BAOLAB/ISHAPE (Larsen 1999) for the cluster image analysis.

The present paper presents the determined structural parameters without
getting into their dynamical or evolutionary interpretation.  The
outline of their importance is given in Hill \& Zaritsky (2006) and will
not be repeated here.  A study of the structural star cluster properties
of the present sample will be reported in a forthcoming paper in
combination with the population analysis.

\sectionb{2}{CLUSTER SAMPLE}

The Suprime-Cam frames employed in this study were taken from the survey
of compact star clusters described in Paper I. This survey was conducted
in the field of $\sim$$17.5\arcmin \times 28.5\arcmin$ size centered at
$0^{\rm h}40\fm9$, $+40\degr45\arcmin$ (J2000.0).  We used the $V$-band
frames ($5\times 2$\,min. exposures) of M\,31 that were secured during
the verification period of Suprime-Cam in July of 1999.  The mosaic
camera consisted of 8 CCDs containing $\sim$$8K \times 8K$ pixels of
$0.2\arcsec \times 0.2\arcsec$ size in total.  The typical full width at
half maximum (FWHM) of star images on different CCDs and in different
exposures is of $\sim$$0.7\arcsec$.  The raw data were processed in a
standard manner and a stacked $V$-band mosaic image was produced.  The
morphological atlases and aperture photometry results for prominent
compact objects ($17.5 \sim< V \sim< 19.5$) were presented in Paper I
separately for 52 H$\alpha$ emission objects (KWE) and 49 non-emission
clusters (KWC).  The signal-to-noise of the faintest studied cluster
(photometric limit of the measured mosaic image $V\sim25.5$) is
satisfactory for the employed profile fitting procedure (Larsen 1999).
Further image reduction, object selection and photometry procedure
details are to be found in Paper I.

For the present study 49 KWC and 2 KWE compact objects were selected
from the corresponding catalogs of Paper I. The KWC object list was
supplemented with two KWE objects satisfying the upper magnitude limit,
$V \sim 19.5$, which was generally applied for the selection of the KWC
objects.  The sample of the clusters selected for the present study is
listed in Table~1.

{\small\tabcolsep=5pt
\begin{longtable}{ccccccr}
\multicolumn {7}{c}{\parbox{90mm}{{\normbf Table~1.}
{\norm Basic parameters of the compact star clusters.}}}\\
\noalign{\smallskip} \hline \noalign{\smallskip}
\multicolumn{1}{c}{Cluster}& \multicolumn{1}{c}{RA(2000)}&
\multicolumn{1}{c}{Dec(2000)}& \multicolumn{1}{c}{$V^*$}&
\multicolumn{1}{c}{$(B-V)^*$}& \multicolumn{1}{c}{$b/a$} &
\multicolumn{1}{c}{PA}\\
\hline
\noalign{\vspace{3pt}}
\endfirsthead
\multicolumn {7}{c}{\parbox{90mm}{{\normbf Table~1.}
{\norm Continued}}}\\
\noalign{\smallskip} \hline \noalign{\smallskip}
\multicolumn{1}{c}{Cluster}& \multicolumn{1}{c}{RA(2000)}&
\multicolumn{1}{c}{Dec(2000)}& \multicolumn{1}{c}{$V^*$}&
\multicolumn{1}{c}{$(B-V)^*$}& \multicolumn{1}{c}{$b/a$} &
\multicolumn{1}{c}{PA}\\
\hline
\noalign{\vspace{3pt}}
\endhead
\noalign{\vspace{3pt}}
\hline
\endfoot
          KWC01  &  10.04577 & 40.60326 & 18.50   &   0.21   &  0.44  &  118   \\
          KWC02  &  10.05932 & 40.65581 & 18.82   &   0.27   &  0.71  &  85    \\
          KWC03  &  10.06074 & 40.62242 & 18.57   &   0.05   &  0.64  &  165   \\
          KWC04  &  10.06408 & 40.61515 & 18.59   &   0.31   &  0.44  &  72    \\
          KWC05  &  10.06460 & 40.66653 & 18.49   &   0.30   &  0.80  &  126   \\
          KWC06  &  10.07201 & 40.65136 & 18.13   &   0.13   &  0.74  &  159   \\
          KWC07  &  10.07320 & 40.65556 & 18.56   &   0.10   &  0.49  &  88    \\
          KWC08  &  10.07620 & 40.54577 & 18.56   &   0.41   &  0.81  &  60    \\
          KWC09  &  10.07855 & 40.52101 & 19.33   &   1.57   &  0.64  &  105   \\
          KWC10  &  10.08099 & 40.62479 & 18.69   &   0.39   &  0.70  &  137   \\
          KWC11  &  10.08302 & 40.51324 & 18.74   &   0.16   &  0.29  &  76    \\
          KWC12  &  10.09631 & 40.51323 & 17.48   &   0.67   &  0.81  &  51    \\
          KWC13  &  10.10304 & 40.63052 & 17.22   &   1.39   &  0.44  &  50    \\
          KWC14  &  10.10351 & 40.81297 & 19.23   &   0.55   &  0.87  &  59    \\
          KWC15  &  10.10370 & 40.81690 & 19.69   &   0.78   &  0.85  &  85    \\
          KWC16  &  10.10807 & 40.62813 & 19.33   &   1.16   &  0.67  &  50    \\
          KWC17  &  10.11374 & 40.75674 & 18.74   &   0.20   &  0.84  &  25    \\
          KWC18  &  10.13581 & 40.83711 & 19.14   &   0.26   &  0.65  &  119   \\
          KWC19  &  10.15227 & 40.67085 & 19.03   &   0.64   &  0.68  &  56    \\
          KWC20  &  10.15519 & 40.65408 & 19.14   &   0.67   &  0.50  &  120   \\
          KWC21  &  10.15569 & 40.81267 & 19.25   &   0.30   &  0.71  &  142   \\
          KWC22  &  10.17370 & 40.64113 & 18.73   &   1.21   &  0.88  &  104   \\
          KWC23  &  10.17619 & 40.60127 & 19.23   &   0.76   &  0.91  &  171   \\
          KWC24  &  10.18421 & 40.74610 & 19.82   &   1.54   &  0.46  &  135   \\
          KWC25  &  10.19348 & 40.86130 & 19.18   &   0.38   &  0.70  &  70    \\
          KWC26  &  10.20151 & 40.58503 & 18.41   &   0.29   &  0.58  &  83    \\
          KWC27  &  10.20159 & 40.86619 & 18.62   &   0.25   &  0.65  &  25    \\
          KWC28  &  10.20241 & 40.96009 & 19.28   &   0.55   &  0.93  &  156   \\
          KWC29  &  10.20591 & 40.69223 & 18.28   &   0.80   &  0.95  &  159   \\
          KWC30  &  10.21459 & 40.55770 & 19.27   &   0.65   &  0.88  &  116   \\
          KWC31  &  10.21512 & 40.73504 & 17.98   &   0.74   &  0.60  &  21    \\
          KWC32  &  10.21786 & 40.89896 & 19.29   &   0.94   &  0.85  &  135   \\
          KWC33  &  10.21782 & 40.97817 & 19.45   &   0.62   &  0.91  &  58    \\
          KWC34  &  10.22069 & 40.58883 & 18.88   &   0.88   &  0.90  &  103   \\
          KWC35  &  10.23960 & 40.74087 & 19.08   &   1.27   &  0.87  &  134   \\
          KWC36  &  10.27868 & 40.57474 & 19.15   &   0.31   &  0.78  &  127   \\
          KWC37  &  10.28315 & 40.88356 & 18.70   &   0.94   &  0.96  &  12    \\
          KWC38  &  10.29172 & 40.96973 & 19.01   &   0.72   &  0.94  &  158   \\
          KWC39  &  10.30333 & 40.57155 & 18.10   &   0.26   &  0.94  &  173   \\
          KWC40  &  10.30768 & 40.56615 & 18.32   &   0.23   &  0.84  &  129   \\
          KWC41  &  10.32568 & 40.73369 & 19.18   &   0.55   &  0.96  &  100   \\
          KWC42  &  10.32810 & 40.95436 & 17.67   &   1.42   &  0.88  &  118   \\
          KWC43  &  10.33723 & 40.98452 & 18.11   &   1.18   &  0.98  &  108   \\
          KWC44  &  10.35044 & 40.61307 & 18.69   &   0.14   &  0.67  &  12    \\
          KWC45  &  10.36251 & 40.69373 & 19.22   &   0.43   &  0.81  &  140   \\
          KWC46  &  10.40363 & 40.79040 & 18.69   &   0.48   &  0.80  &  139   \\
          KWC47  &  10.40855 & 40.56967 & 18.95   &   0.61   &  0.35  &  33    \\
          KWC48  &  10.41051 & 40.82676 & 19.29   &   0.51   &  0.89  &  106   \\
          KWC49  &  10.41184 & 40.68182 & 18.11   &   0.27   &  0.84  &  98    \\
          KWE33  &  10.41127 & 40.73314 & 18.79   &   0.02   &  0.86  &  96    \\
          KWE52  &  10.19922 & 40.98502 & 18.46   &   0.39   &  0.48  &  75    \\
\end{longtable}
}
{\bf Notes to Table 1:} * -- photometry from Narbutis et al. (2006); $b/a$ --
minor to major axis ratio; PA -- major axis position angle, in degrees
from North to East.

{\small \tabcolsep=3pt
\begin{longtable}{lcccccccccc}
\multicolumn {11}{c}{\parbox{110mm}{ {\normbf Table~2.}
{\norm Structural parameters of the compact star clusters.}}}\\
\noalign{\smallskip} \hline \noalign{\smallskip}
\multicolumn{1}{c}{Cluster}& \multicolumn{1}{c}{W$_{\rm EFF}$}&
\multicolumn{1}{c}{$\sigma$(W$_{\rm EFF}$)}&
\multicolumn{1}{c}{$\gamma$}& \multicolumn{1}{c}{$\gamma_{\rm
min}$}& \multicolumn{1}{c}{$\gamma_{\rm max}$}&
\multicolumn{1}{c}{W$_{\rm King}$}&
\multicolumn{1}{c}{$\sigma$(W$_{\rm King}$)}&
\multicolumn{1}{c}{$c$}& \multicolumn{1}{c}{$c_{\rm min}$}&
\multicolumn{1}{c}{$c_{\rm max}$}\\
\hline
\noalign{\vspace{3pt}}
\endfirsthead
\multicolumn {11}{c}{\parbox{110mm}{ {\normbf Table~2.}
{\norm Continued}}}\\
\noalign{\smallskip} \hline \noalign{\smallskip}
\multicolumn{1}{c}{Cluster}& \multicolumn{1}{c}{W$_{\rm EFF}$}&
\multicolumn{1}{c}{$\sigma$(W$_{\rm EFF}$)}&
\multicolumn{1}{c}{$\gamma$}& \multicolumn{1}{c}{$\gamma_{\rm
min}$}& \multicolumn{1}{c}{$\gamma_{\rm max}$}&
\multicolumn{1}{c}{W$_{\rm King}$}&
\multicolumn{1}{c}{$\sigma$(W$_{\rm King}$)}&
\multicolumn{1}{c}{$c$}& \multicolumn{1}{c}{$c_{\rm min}$}&
\multicolumn{1}{c}{$c_{\rm max}$}\\
\hline
\noalign{\vspace{3pt}}
\endhead
\noalign{\vspace{3pt}}
\hline
\endfoot
         KWC01 &   0.54   & 0.02 &  5.2    &  5.0   &  5.4   &   0.30  &   0.06 &    1.4 &     1.2 &     1.4   \\
         KWC02 &   0.22   & 0.04 &  2.0    &  1.8   &  2.2   &   0.24  &   0.06 &    2.0 &     1.5 &     3.0   \\
         KWC03 &   0.58   & 0.04 &  4.2    &  4.0   &  4.4   &   0.56  &   0.04 &    0.8 &     0.7 &     0.8   \\
         KWC04 &   1.04   & 0.06 &  10.    &  10.   &  70.   &   0.80  &   0.10 &    0.5 &     0.4 &     0.7   \\
         KWC05 &   0.50   & 0.04 &  4.4    &  3.8   &  5.0   &   0.32  &   0.02 &    1.2 &     1.2 &     1.2   \\
         KWC06 &   1.22   & 0.04 & $>$20   &  --    &  --    &   1.24  &   0.06 &    0.1 &     0.1 &     0.7   \\
         KWC07 &   0.54   & 0.04 &  3.4    &  3.0   &  3.8   &   0.40  &   0.04 &    1.3 &     1.2 &     1.3   \\
         KWC08 &   1.74   & 0.04 &  4.8    &  3.6   &  5.0   &   1.60  &   0.06 &    0.7 &     0.7 &     1.0   \\
         KWC09 &   0.22   & 0.04 &  1.8    &  1.6   &  2.0   &   0.34  &   0.04 &    2.0 &     1.8 &     3.0   \\
         KWC10 &   0.48   & 0.04 &  2.2    &  2.0   &  2.4   &   0.46  &   0.04 &    2.0 &     1.4 &     3.2   \\
         KWC11 &   0.88   & 0.04 & $>$20   &  --    &  --    &   0.86  &   0.04 &    0.3 &     0.2 &     0.3   \\
         KWC12 &   13.0   & 2.00 &  3.0    &  1.4   &  6.0   &   16.2  &   2.00 &    1.5 &     0.5 &     2.0   \\
         KWC13 &   0.40   & 0.06 &  1.4    &  1.2   &  1.4   &   0.92  &   0.04 &    2.6 &     2.3 &     2.8   \\
         KWC14 &   2.70   & 0.10 &  3.8    &  2.4   &  4.0   &   2.72  &   0.10 &    1.5 &     0.7 &     2.0   \\
         KWC15 &   1.22   & 0.06 &  2.4    &  2.0   &  2.6   &   1.20  &   0.06 &    1.5 &     0.7 &     2.0   \\
         KWC16 &   0.14   & 0.04 &  0.6    &  0.6   &  0.8   &   1.58  &   0.06 &    2.0 &     1.7 &     2.3   \\
         KWC17 &   0.40   & 0.04 &  2.6    &  2.4   &  2.8   &   0.28  &   0.04 &    1.6 &     1.5 &     1.8   \\
         KWC18 &   0.50   & 0.06 &  2.4    &  2.2   &  2.6   &   0.44  &   0.06 &    1.5 &     1.2 &     1.7   \\
         KWC19 &   0.76   & 0.04 &  2.8    &  2.6   &  3.0   &   0.66  &   0.06 &    1.4 &     1.2 &     1.5   \\
         KWC20 &   1.50   & 0.04 & $>$20   &  --    &  --    &   1.08  &   0.06 &    0.1 &     0.0 &     0.3   \\
         KWC21 &   0.74   & 0.04 &  2.0    &  1.8   &  2.2   &   0.78  &   0.04 &    1.7 &     1.2 &     2.4   \\
         KWC22 &   1.80   & 0.04 &  3.0    &  2.6   &  3.6   &   1.74  &   0.04 &    1.2 &     0.6 &     2.0   \\
         KWC23 &   1.48   & 0.06 &  1.8    &  1.6   &  2.0   &   1.52  &   0.04 &    1.5 &     1.2 &     2.1   \\
         KWC24 &   0.30   & 0.06 &  1.0    &  0.8   &  1.2   &   1.34  &   0.06 &    2.2 &     1.5 &     2.2   \\
         KWC25 &   0.96   & 0.04 &  16.    &  6.8   &  24.   &   0.90  &   0.08 &    0.3 &     0.1 &     0.7   \\
         KWC26 &   1.00   & 0.04 &  14.    &  13.   &  23.   &   1.00  &   0.12 &    0.3 &     0.2 &     0.6   \\
         KWC27 &   0.34   & 0.04 &  2.8    &  2.6   &  3.0   &   0.18  &   0.04 &    2.0 &     1.5 &     3.4   \\
         KWC28 &   0.64   & 0.02 &  3.2    &  3.0   &  3.4   &   0.54  &   0.06 &    1.0 &     1.0 &     1.2   \\
         KWC29 &   0.80   & 0.04 &  2.0    &  2.0   &  2.0   &   0.82  &   0.04 &    1.5 &     1.4 &     2.3   \\
         KWC30 &   2.66   & 0.04 &  4.8    &  2.4   &  8.8   &   2.64  &   0.06 &    1.1 &     0.5 &     1.5   \\
         KWC31 &   0.78   & 0.06 &  5.0    &  4.8   &  8.2   &   0.68  &   0.06 &    0.7 &     0.7 &     1.1   \\
         KWC32 &   1.86   & 0.06 &  2.4    &  2.0   &  4.0   &   1.82  &   0.02 &    1.2 &     0.7 &     2.0   \\
         KWC33 &   0.50   & 0.02 &  3.0    &  2.8   &  3.0   &   0.40  &   0.02 &    1.2 &     1.2 &     1.3   \\
         KWC34 &   2.82   & 0.20 & $>$20   &  --    &  --    &   3.00  &   0.40 &    0.7 &     0.4 &     1.1   \\
         KWC35 &   0.28   & 0.02 &  2.0    &  1.8   &  2.0   &   0.30  &   0.02 &    2.4 &     2.0 &     2.9   \\
         KWC36 &   1.10   & 0.02 & $>$20   &  --    &  --    &   0.86  &   0.06 &    0.5 &     0.5 &     0.7   \\
         KWC37 &   0.38   & 0.02 &  1.8    &  1.6   &  1.8   &   0.48  &   0.04 &    1.9 &     1.7 &     2.3   \\
         KWC38 &   1.20   & 0.02 &  3.0    &  2.8   &  3.2   &   1.06  &   0.06 &    1.2 &     0.8 &     2.0   \\
         KWC39 &   0.74   & 0.08 &  4.4    &  3.8   &  5.0   &   0.56  &   0.04 &    1.2 &     0.7 &     1.3   \\
         KWC40 &   1.70   & 0.06 &  2.0    &  1.8   &  2.4   &   1.70  &   0.06 &    2.0 &     1.5 &     2.1   \\
         KWC41 &   0.98   & 0.02 &  2.2    &  2.0   &  2.4   &   0.96  &   0.02 &    1.5 &     1.2 &     2.3   \\
         KWC42 &   0.48   & 0.02 &  2.4    &  2.4   &  2.4   &   0.44  &   0.04 &    1.4 &     1.3 &     1.5   \\
         KWC43 &   0.32   & 0.02 &  2.0    &  2.0   &  2.0   &   0.34  &   0.02 &    2.0 &     1.8 &     2.8   \\
         KWC44 &   3.04   & 0.20 &  5.0    &  3.6   &  9.4   &   3.00  &   0.20 &    0.7 &     0.6 &     0.9   \\
         KWC45 &   0.54   & 0.04 &  3.2    &  2.8   &  4.8   &   0.46  &   0.08 &    1.0 &     0.7 &     1.2   \\
         KWC46 &   2.08   & 0.06 &  4.0    &  3.0   &  5.4   &   2.06  &   0.06 &    1.0 &     0.7 &     1.5   \\
         KWC47 &   0.74   & 0.04 &  3.0    &  2.8   &  3.2   &   0.60  &   0.06 &    1.3 &     1.2 &     1.5   \\
         KWC48 &   0.52   & 0.04 &  4.4    &  4.0   &  5.0   &   0.40  &   0.08 &    1.0 &     0.7 &     1.2   \\
         KWC49 &   0.36   & 0.04 &  3.0    &  2.8   &  3.6   &   0.30  &   0.04 &    1.3 &     1.2 &     1.5   \\
         KWE33 &   0.44   & 0.04 &  2.0    &  1.8   &  2.0   &   0.48  &   0.04 &    2.3 &     1.5 &     2.8   \\
         KWE52 &   0.14   & 0.06 &  2.2    &  2.0   &  2.4   &   0.10  &   0.04 &    2.7 &     2.0 &     3.6   \\
\end{longtable}
}

{\bf Notes to Table 2:} W$_{\rm EFF}$, $\sigma$(W$_{\rm EFF}$) -- FWHM
from the EFF model fit and its r.m.s., in arcseconds; $\gamma$,
$\gamma_{\rm min}$, $\gamma_{\rm max}$ -- the EFF model power-law index
and its lower and upper limits, respectively; W$_{\rm King}$,
$\sigma$(W$_{\rm King}$) -- FWHM from the King model fit and its r.m.s.,
in arcseconds; $c$, $c_{\rm min}$, $c_{\rm max}$ -- the King model
concentration parameter, $c$\,=\,log$(r_t/r_c)$, and its lower and upper
limits, respectively.  W notation for FWHM was applied to shorten
headline of Table~2.

\vspace{8mm}

{\it UBVRI} broad-band aperture CCD photometry for these clusters,
performed on the Local Group Galaxy Survey mosaic images (Massey et al.
2006), was recently published by Narbutis et al.  (2006). The
$V$-band magnitudes and $B-V$ colors from this paper are repeated in
Table 1 for convenience.  In the present study we adopt the M\,31
distance modulus of $m - M = 24.5$ (e.g., Stanek \& Garnavich 1998;
Holland 1998).

\sectionb{3}{CLUSTER STRUCTURAL PARAMETERS}

\subsectionb{3.1}{Profile fitting models}

Two different analytical models -- King (1962, 1966) and EFF (Elson et
al. 1987) -- were used for the determination of star cluster structural
parameters.  The King model was empirically derived to reproduce surface
brightness profiles of the Milky Way (MW) globular clusters and later
was deduced from tidally limited models of isothermal spheres (King
1966).  Therefore, they are expected to describe bounded and dynamically
relaxed stellar systems possessing isothermal and isotropic star
distribution functions (stars are of the same mass and reside within a
tidal field exerted by another object).  The King model is defined by
the three parameters -- central surface brightness, $\mu_0$, core
radius, $r_c$, and tidal radius, $r_t$:

$$
\mu(r) = \mu_{0} \left[ {\left(\displaystyle1+\displaystyle{r^{2}
\over \displaystyle
r_{c}^{2}}\right)^{-{1\over2}}}-{\left(\displaystyle
1+{\displaystyle r_{t}^{2} \over \displaystyle
r_{c}^{2}}\right)^{-{1\over2}}} \right]^2 \eqno(1)
$$

\noindent The EFF analytical model was empirically derived by Elson et
al.  (1987) to reproduce surface brightness profiles of young ($<$300
Myr) star clusters in the Large Magellanic Cloud.  This model is
expected to trace the stellar systems, which are still actively loosing
stars due to the tidal field they reside in.  Absence of a
well-established tidal limit is interpreted as  dynamical youth and it
is the main difference from dynamically older systems described by the
King models.  The EFF model is described by a combination of the three
parameters -- central surface brightness, $\mu_0$, scale length,
$r_{e}$, and power-law index, $\gamma$:

$$
\mu(r) = \mu_0 \left( 1 + {r^{2} \over r_{e}^{2}} \right)^{-{\gamma
\over 2}} \eqno(2)
$$

\noindent The EFF model parameter $r_{e}$ represents the scale-length
rather than the core radius, and the latter can be derived as $r_c =
r_{e} \sqrt{2^{2/\gamma}-1}$ (Elson et al. 1987).

\subsectionb{3.2}{Fitting procedure}

The structural parameters of the clusters were derived by fitting 2-D
King
and EFF (Elson et al. 1987) models to the $V$-band cluster images with
the BAOLAB/ISHAPE program package (Larsen 1999).  The ISHAPE procedure
convolves a 2-D analytical model with a Point Spread Function (PSF),
constructed from well isolated stars in the field, and compares it with
the observed cluster image using a reduced $\chi^2$ test by iteratively
adjusting parameters of the analytical models until the best fit to the
data is obtained.  A PSF for the ISHAPE procedure was generated with the
{\it seepsf} task from the DAOPHOT (Stetson 1987), which is implemented
in the IRAF software package (Tody 1993).

We used the coordinates given in Paper I to extract individual images
for each cluster of $20\arcsec \times 20\arcsec$ size suitable for
fitting with the ISHAPE procedure.  Our cluster sample is dominated by
semi-resolved objects, which cannot be fitted with the ISHAPE procedure
by allowing a free fitting of cluster centers.  Therefore, as the first
step we determined accurate coordinates of the cluster centers using
histogram plots constructed by integrating cluster images along R.A. and
Dec. coordinates.  To avoid spurious results due to contamination by
neighboring sources, we constrained the histogram plots to the area of
$4\arcsec \times 4\arcsec$ ($21 \times 21$ pixels) around the
approximate cluster centers.  The derived object centers were fixed and
used in the ISHAPE procedure.

Fitting was performed running the ISHAPE procedure with a variable
fitting radius from $1.4\arcsec$ to $2.2\arcsec$ depending on the actual
size of the cluster.  The lower limit of the fitting radius was set to
be approximately equal to 2$\cdot$FWHM of the PSF (FWHM of the stellar
images on $V$-band frames is equal to $\sim$$0.7\arcsec$), thus allowing
us to account for the influence of the PSF and to measure the intrinsic
star cluster parameters reliably.

The EFF model was fit six times:  with a fixed power-law index, $\gamma
= 2$, 3, 5, and allowing a power-law index to vary as a free parameter
starting from different initial guesses, ${\gamma_0} = 2$, 3 and 5. The
King model was fit eight times:  with a fixed concentration factor, $r_t
/{r_c} = 5$, 15, 30 or 100, and allowing $r_t /{r_c}$ to vary as a free
parameter starting from different initial guesses, $(r_t /{r_c})_0 = 5$,
15, 30 or 100.  The derived structural parameters for both models are
presented in Tables 1 and 2, and briefly discussed in the next section.

\begin{center}
\vbox{\small \tabcolsep=5pt
\begin{tabular}  {clccccc}
\multicolumn {7}{l}{\parbox{10cm}{ {\normbf \ \ Table~3.} {\norm\
The cluster structural parameters derived from the HST and Subaru data.}}}\\
\tablerule \multicolumn{1}{c}{Cluster}& \multicolumn{1}{l}{Source}&
\multicolumn{1}{c}{$r_c$}& \multicolumn{1}{c}{$r_t$}& \multicolumn{1}{c}{$c$} &
\multicolumn{1}{c}{$b/a$}& \multicolumn{1}{c}{PA} \\
\tablerule
KWC29 & Barmby et al. (2002) & 0.37 & 17.42 & 1.68 & 0.85$\pm$0.04 & 4$\pm$2  \\
      & This study           & 0.42 & 12.65 & 1.48 & 0.95$\pm$0.05 & 159$\pm$30 \\
\hline
KWC42 & Barmby et al. (2002) & 0.21 &  5.50 & 1.42 & 0.90$\pm$0.01 & 118$\pm$17 \\
      & This study           & 0.23 &  5.69 & 1.40 & 0.88$\pm$0.02 & 118$\pm$15 \\
\hline
\end{tabular}
}
\end{center}

{\bf Notes to Table 3:} $r_c$, $r_t$ -- core and tidal radii, in
arcseconds; $c$ -- concentration parameter; $b/a$ --
minor to major axis ratio; PA -- major axis
position angle, in degrees from North to East.

The reliability of our results can be deduced by comparing with previous
studies, which employed high resolution imaging.  Two clusters from our
catalog -- KWC29 and KWC42 -- have estimates of the structural
parameters based on the Hubble Space Telescope (HST) measurements
(Barmby et al. 2002).  A comparison of the derived King model parameters
is presented in Table~3.  The structural cluster parameters from our
study and from Barmby et al.  (2002) for both clusters in common agree
very well.  In fact, the differences of all structural cluster
parameters fall within the range of the parameter uncertainties declared
in both studies.

It is important to note, that the ISHAPE algorithm is designed to work
best with the objects whose intrinsic size is comparable to or even
smaller than FWHM of the image PSF.  Therefore, it could not work well
with large and/or semi-resolved clusters.  Perhaps it would be more
appropriate to fit one-dimensional surface brightness profiles for such
objects.  This method and comparison with the results presented in this
study will be discussed in detail for the same cluster sample in the
forthcoming paper of the M\,31 cluster study series
(\v{S}ablevi\v{c}i\={u}t\.{e} et al. 2007, in preparation).

\sectionb{4}{RESULTS}

Minor to major axis ratios, $b/a$, and major axis position angles, PA,
of the clusters, fitted with the EFF and King models, were averaged and
are given in Table~1.  These parameters, derived basing on different
models, agree well and can be reliably averaged, see Figures 1 and 2. It
is worth noting that the axis ratio of our objects extends far below
0.8, which is the typical lower range of globular clusters, and PA
distributes over a full range of 0\degr--180\degr, although the observed
region is restricted to the S-W field of the M\,31 disk.

The best-fit structural parameters derived with the ISHAPE procedure for
the cases of the EFF and King models are presented in Table~2.  We
provide (Table~2) only parameters directly derived in the ISHAPE
procedure.  FWHM and their accuracy are given for both models.
Power-law index, $\gamma$ (EFF), and concentration parameter,
$c$\,=\,log$(r_t/r_c)$ (King), are provided together with estimated
lower and upper parameter limits.  Figure 3 shows the comparison between
the cluster FWHM obtained using the best-fit EFF and King models.  A
rather tight correlation between FWHM derived basing on independent fits
of two different models suggests that derived parameters are reliable.
Three deviating objects, which are assigned with larger FWHM by fitting
the King model, basing on their images are suspected to be background
galaxy candidates (KWC13, KWC16, and KWC24).  One object with larger
FWHM, obtained by fitting the EFF model, is a suspected asterism
(KWC20).  The largest object in our sample, KWC12, is omitted from
Figures\,3-8 for analysis convenience and due to lower reliability of
the derived parameters, because the cluster's size exceeds the used
fitting radius.

Inter-correlations of the derived parameter are shown in Figures 4--6.
The limit of $\gamma = 9$ was artificially assigned for larger $\gamma$
values in order to make figures more comprehensible.  However, it is
worth noting that such large $\gamma$ values can indicate

\vbox{ \centerline{\psfig{figure=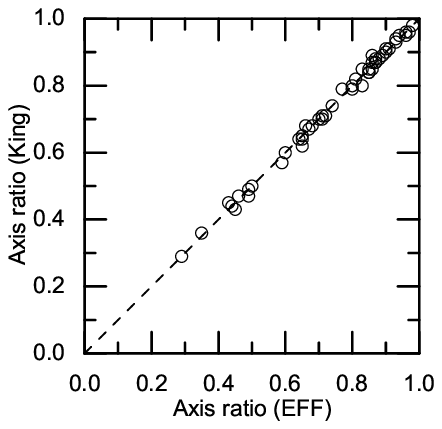,width=84mm,clip=}}
\vspace{-.5mm} \captionb{1}{Minor to major axis ratios, $b/a$, of
the clusters obtained by fitting the EFF and King
models. The dashed line marks a bisector.} %\vspace{3mm}

\vskip5mm \centerline{\psfig{figure=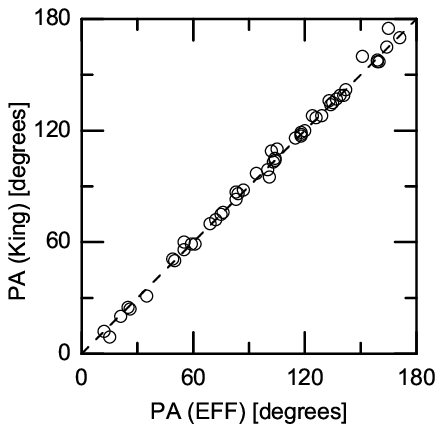,width=84mm,clip=}}
\vspace{-.5mm} \captionb{2}{Major axis position angles of the
clusters obtained by fitting the EFF and King models, in degrees
from North
to East. The dashed line marks a bisector.} %\vspace{3mm}
}

\vbox{ \centerline{\psfig{figure=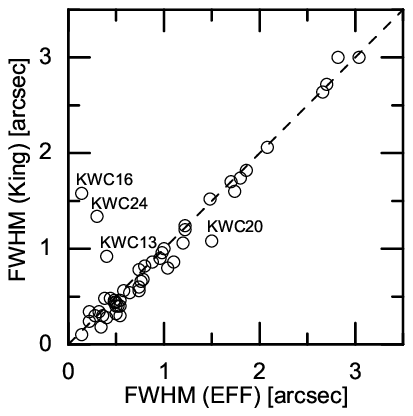,width=84mm,clip=}}
\vspace{-.5mm} \captionb{3}{FWHM of the clusters obtained by
fitting the EFF and King models. The most deviating objects are
indicated. The dashed line marks a bisector.} %\vspace{3mm}
\vskip5mm
\centerline{\psfig{figure=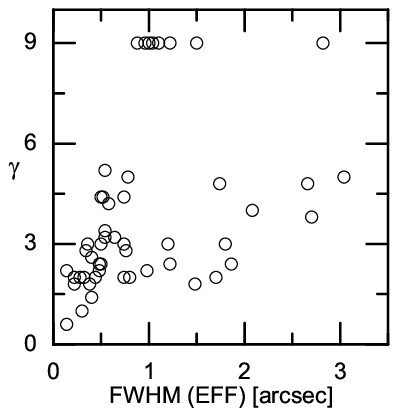,width=84mm,clip=}}
\vspace{-.5mm} \captionb{4}{The EFF model power-law index,
$\gamma$, versus FWHM obtained by fitting the EFF model. The limit
of $\gamma = 9$ was artificially assigned for larger $\gamma$
values.} %\vspace{3mm}
}

\vbox{ \centerline{\psfig{figure=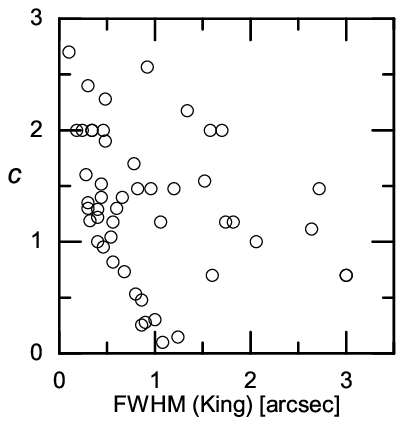,width=84mm,clip=}}
\vspace{-.5mm} \captionb{5}{The King model concentration
parameter, $c$\,=\,log$(r_t/r_c)$, versus FWHM obtained by fitting
the King model.} %\vspace{3mm}

\vskip5mm \centerline{\psfig{figure=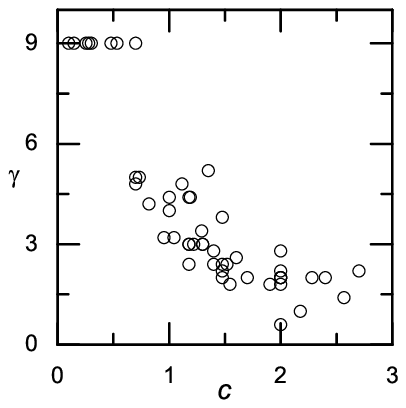,width=84mm,clip=}}
\vspace{-.5mm} \captionb{6}{The EFF model power-law index,
$\gamma$, versus the King model concentration parameter, $c$.} %\vspace{3mm}
}

\vbox{
\centerline{\psfig{figure=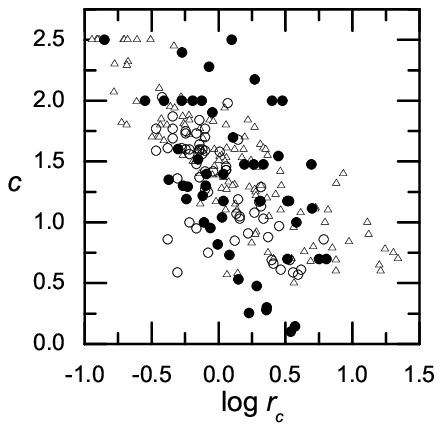,width=80mm,angle=0,clip=}}
\vspace{-1mm} \captionb{7}{The King model cluster structural
parameters ($r_c$ measured in pc). Filled circles mark the M\,31
clusters from this study; open circles mark the M\,31 clusters
from Barmby et al. (2002); open triangles mark the MW globular
clusters from Harris (1996).} %\vspace{3mm}

\vskip3mm
\centerline{\psfig{figure=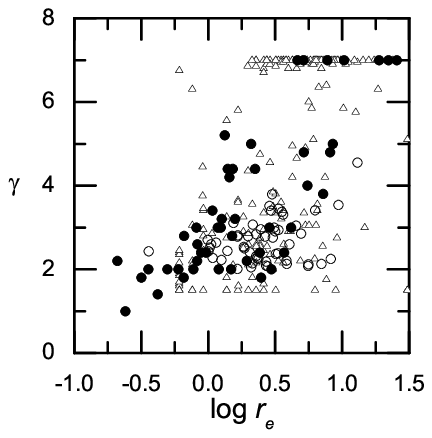,width=80mm,angle=0,clip=}}
\vspace{-1mm} \captionb{8}{The EFF model cluster structural
parameters ($r_e$ measured in pc). Filled circles mark the M\,31
clusters from this study; open circles mark the LMC clusters from
Mackey and Gilmore (2003); open triangles mark the SMC clusters
from Hill and Zaritsky (2006).} %\vspace{3mm}
}

\noindent the object's stellarity or irregular features arising, e.g.,
due to projected bright foreground or background stars.  Figure\,6 shows
good correlation between $\gamma$ and $c$ parameters, thus confirming
that large $\gamma$ values are not fitting errors, but an inherent
property of object images.

Our sample of compact star clusters avoids most of the well-established
globular clusters in M\,31 (they were over-exposed on Suprime-Cam frames),
as well as faint star clusters due to the $V$-band upper magnitude limit
($V \sim 19.5$) applied for selection. Therefore, one should keep in
mind that comparisons, presented below, with the results of previous
star cluster structural parameter studies in M\,31 (Barmby et al. 2002),
MW (Harris 1996), LMC (Mackey \& Gilmore 2003) and SMC (Hill and
Zaritsky 2006) are aimed only to demonstrate general trends without
going into discussion, because of mentioned selection effects inherent
to our sample. Actually, MW sample of non-globular, compact Galaxy disk
clusters would be the most desirable to be compared, however, it is not
representative enough due to observation difficulties trough rather
opaque disk interstellar media, which prevents discovery and study
of the distant low mass young star clusters.

The King model structural parameters of the M\,31 star clusters, derived
in the present study and presented by Barmby et al.  (2002), are
compared with the MW globular cluster parameters (Harris 1996) in Figure
7. The results from these studies are in reasonable agreement.  It can
be deduced from Figure 7, that clusters in both galaxies follow
essentially the same trends.  Objects with $c\leq0.3$ are suspected to
be asterisms.  The galaxy candidates (KWC13, KWC16 and KWC24) all belong
to the object group with slightly larger $r_c$ values, than the majority
of clusters at $c\geq2$.  Note the artificial upper limit of $c = 2.5$,
which was set for our objects in accordance to the data provided by
Harris (1996) for the MW globular clusters.  The parameters, $r_c$ and
$r_e$, used in Figures 7--8, were computed from FWHM, $\gamma$ and $c$
parameters, presented in Table~2, basing on transformation equations (6,
10) provided by Larsen (2006).

The structural parameters of the EFF model for the M\,31 clusters,
derived in the present study, are presented in Figure 8 and compared
with the LMC rich clusters from Mackey \& Gilmore (2003) and the SMC
clusters from Hill \& Zaritsky (2006). In general the M\,31 clusters
from the present study occupy structural parameter space, that is well
overlapping with the Magellanic Clouds cluster parameter ranges. The
galaxy candidates (KWC13, KWC16 and KWC24) again stand out in Figure 8
with $\gamma \leq 1.5$. Note the artificial upper limit of $\gamma = 7$,
which was set for our objects in accordance to the data provided by
Hill \& Zaritsky (2006).

An evolutionary analysis of the structural and photometric parameters
of the M\,31 clusters will be performed in the forthcoming paper
(Vansevi{\v c}ius et al. 2007, in preparation).

\sectionb{5}{SUMMARY}

We present the structural parameters for 51 compact star clusters from
the survey of compact star clusters in the South-West field of the
M\,31 disk by Kodaira et al. (2004). The 2-D image fitting
procedure ISHAPE (Larsen 1999) was employed for determination of the
cluster parameters. The structural parameters were derived on the
$V$-band images using the King and EFF analytical models.

The structural parameters of two clusters derived in this study are in
very good agreement with the earlier estimates based on the HST data by
Barmby et al.  (2002), see Table~3.  In general, the M\,31 compact star
cluster King model's structural parameters overlap with the parameter
region of the MW globular clusters.  The M\,31 star cluster parameters
span the EFF model structural parameter range similar to the available
data for SMC star clusters, but existing subtle differences are due for
further study.

\vskip 5mm

\thanks{We are thankful to Valdas Vansevi{\v c}ius for correcting
the manuscript.  This study was financially supported in part by a
Grant T-08/06 of the Lithuanian State Science and Studies
Foundation. This research has made use of SAOImage DS9, developed by
Smithsonian Astrophysical Observatory.}

\References

\enlargethispage{3mm}

\refb Barmby~P., Holland~S., Huchra~J. 2002, AJ, 123, 1937

\refb Bastian~N., Gieles~M., Efremov~Yu.~N., Lamers~H. 2005, A\&A,
443, 79

\refb Beasley~M.~A., Brodie~J.~P., Strader~J., Forbes~D.~A.,
Proctor~R.~N., Barmby~P., Huchra~J.~P. 2004, AJ, 128, 1623

\refb Beasley~M.~A., Brodie~J.~P., Strader~J., Forbes~D.~A.,
Proctor~R.~N., Barmby~P., Huchra~J.~P. 2005, AJ, 129, 1412

\refb Burstein~D. et al. 2004, ApJ, 614, 158

\refb Efremov~Yu.~N., Elmegreen~B.G. 1998, MNRAS, 299, 588

\refb Elson~R.~A.~W., Fall~S.~M., Freeman~K.~C. 1987, ApJ, 323, 54

\refb Fan~Z., Ma~J., de Grijs~R., Yang~Y., Zhou~X. 2006, MNRAS,
371, 1648

\refb Fusi Pecci~F., Bellazzini~M., Buzzoni~A., De Simone~E.,
Federici~L., Galleti~S. 2005, AJ, 130, 554

\refb Galleti~S., Federici~L., Bellazzini~M., Buzzoni~A., Fusi
Pecci~F. 2006, A\&A, 456, 985

\refb Galleti~S., Federici~L., Bellazzini~M., Fusi Pecci~F.,
Macrina~S. 2004, A\&A, 416, 917

\refb Harris~W.~E. 1996, AJ, 112, 1487

\refb Hill~A., Zaritsky~D. 2006, AJ, 131, 414

\refb Holland~S. 1998, AJ, 115, 1916

\refb King~I. 1962, AJ, 67, 471

\refb King~I. 1966, AJ, 71, 64

\refb Kodaira~K., Vansevi\v{c}ius~V., Brid\v{z}ius~A.,
Komiyama~Y., Miyazaki~S., Stonkut\.{e}~R.,
\v{S}ablevi\v{c}i\={u}t\.{e}~I., Narbutis~D. 2004, PASJ, 56, 1025
(Paper I)

\refb Larsen~S.~S. 1999, A\&AS, 139, 393

\refb Larsen~S.~S. 2006, An ISHAPE user's guide (ver. Oct. 23,
2006; available at http://www.astro.uu.nl/$\sim$larsen/baolab/),
p. 14

\refb Lee~M.~G., Chandar~R., Whitmore~B.~C. 2005, AJ, 130, 2128

\refb Mackey~A.~D., Gilmore~G.~F. 2003, MNRAS, 338, 85

\refb Massey~P., Olsen~K.~A.~G., Hodge~P.~W., Strong~S.~B.,
Jacoby~G.~H., Schlingman~W., Smith~R.~C. 2006, AJ, 131, 2478

\refb Miyazaki~S. et al. 2002, PASJ, 54, 833

\refb Morrison~H.~L., Harding~P., Perrett~K., Hurley-Keller~D.
2004, ApJ, 603, 87

\refb Narbutis~D., Vansevi\v{c}ius~V., Kodaira~K.,
\v{S}ablevi\v{c}i\={u}t\.{e}~I., Stonkut\.{e}~R., Brid\v{z}ius~A.
2006, Baltic Astronomy, 15, 461

\refb Rafelski~M., Zaritsky~D. 2005, AJ, 129, 2701

\refb Stanek~K.~Z., Garnavich~P.~M. 1998, ApJ, 503, L131

\refb Stetson~P.~B. 1987, PASP, 99, 191

\refb Tody~D. 1993, in {\it Astronomical Data Analysis Software
and Systems II}, eds. R.~J.~Hanisch, R.~J.~B.~Brissended,
J.~Barnes, ASP Conf. Ser. 52, ASP, San Francisco, 173

\refb Williams~B.~F., Hodge~P.~W. 2001a, ApJ, 548, 190

\refb Williams~B.~F., Hodge~P.~W. 2001b, ApJ, 559, 851

\end{document}